\documentclass[twocolumn]{aastex631}
\usepackage[T1]{fontenc}
\usepackage[utf8]{inputenc}
\let\tablenum\relax
\usepackage{siunitx}
\usepackage{physics}
\usepackage{graphicx}
\usepackage{url}
\usepackage{comment}

\graphicspath{{./}{figures_v1/}}

\begin{document}

\title{A hide-and-seek game: Looking for Population III stars during the Epoch of Reionization through the HeII$\lambda$1640 line}

\correspondingauthor{Alessandra Venditti}
\email{alessandra.venditti@inaf.it}

\author[0000-0003-2237-0777]{Alessandra Venditti}
\affiliation{Dipartimento di Fisica, Sapienza, Universit$\grave{a}$ di Roma, Piazzale Aldo Moro 5, 00185, Roma, Italy}
\affiliation{INFN, Sezione di Roma I, Piazzale Aldo Moro 2, 00185, Roma, Italy}
\affiliation{INAF-Osservatorio Astronomico di Roma, Via di Frascati 33, 00078, Monte Porzio Catone, Italy}
\affiliation{Dipartimento di Fisica, Tor Vergata, Universit$\grave{a}$ di Roma, Via Cracovia 50, 00133, Roma, Italy}

\author[0000-0003-0212-2979]{Volker Bromm}
\affiliation{Department of Astronomy, University of Texas at Austin, 2515 Speedway, Stop C1400, Austin, TX 78712, USA}
\affiliation{Weinberg Institute for Theoretical Physics, University of Texas at Austin, Austin, TX 78712, USA}

\author[0000-0001-8519-1130]{Steven L. Finkelstein}
\affiliation{Department of Astronomy, University of Texas at Austin, 2515 Speedway, Stop C1400, Austin, TX 78712, USA}

\author[0000-0003-2536-1614]{Antonello Calabrò}
\affiliation{INAF-Osservatorio Astronomico di Roma, Via di Frascati 33, 00078, Monte Porzio Catone, Italy}

\author[0000-0002-8951-4408]{Lorenzo Napolitano}
\affiliation{INAF-Osservatorio Astronomico di Roma, Via di Frascati 33, 00078, Monte Porzio Catone, Italy}
\affiliation{Dipartimento di Fisica, Sapienza, Universit$\grave{a}$ di Roma, Piazzale Aldo Moro 5, 00185, Roma, Italy}

\author[0000-0002-9231-1505]{Luca Graziani}
\affiliation{Dipartimento di Fisica, Sapienza, Universit$\grave{a}$ di Roma, Piazzale Aldo Moro 5, 00185, Roma, Italy}
\affiliation{INFN, Sezione di Roma I, Piazzale Aldo Moro 2, 00185, Roma, Italy}
\affiliation{INAF-Osservatorio Astronomico di Roma, Via di Frascati 33, 00078, Monte Porzio Catone, Italy}

\author[0000-0001-9317-2888]{Raffaella Schneider}
\affiliation{Dipartimento di Fisica, Sapienza, Universit$\grave{a}$ di Roma, Piazzale Aldo Moro 5, 00185, Roma, Italy}
\affiliation{INFN, Sezione di Roma I, Piazzale Aldo Moro 2, 00185, Roma, Italy}
\affiliation{INAF-Osservatorio Astronomico di Roma, Via di Frascati 33, 00078, Monte Porzio Catone, Italy}



\begin{abstract}
The gas surrounding first-generation (Pop III) stars is expected to emit a distinct signature in the form of the HeII recombination line at 1640~\AA (HeII$\lambda$1640). Here we explore the challenges and opportunities in identifying this elusive stellar population via the HeII$\lambda$1640 in $M_\star > 10^{7.5} ~ \mathrm{M_\odot}$ galaxies during the Epoch of Reionization (EoR, $z \simeq 6 - 10$), using JWST/NIRSpec. With this aim in mind, we combine cosmological \texttt{dustyGadget} simulations with analytical modeling of the intrinsic HeII emission. While tentative candidates with bright HeII emission like GN-z11 have been proposed in the literature, the prevalence of such bright systems remains unclear due to significant uncertainties involved in the prediction of the HeII luminosity. In fact, similar Pop III clumps might be almost two orders of magnitude fainter, primarily depending on the assumed Pop III-formation efficiency and initial mass function in star-forming clouds, while the effect of stellar mass loss is responsible for a factor of order unity. Moreover, up to $\sim 90 \%$ of these clumps might be missed with NIRSpec/MOS due to the limited FoV, while this problem appears to be less severe with NIRSpec/IFU. We investigate the potential of deep spectroscopy targeting peripheral Pop III clumps around bright, massive galaxies to achieve a clear detection of the first stars.
\end{abstract}



\section{Introduction} 
\label{sec:intro}

The deployment of JWST has opened new frontiers for modern astrophysics, enabling us to explore the depth of the high-$z$ Universe with unprecedented sensitivity and resolution. Notably, it paves the way for the exciting possibility of directly detecting the first generation of stars, known as Population III (Pop\ III) stars.

The HeII recombination line at 1640~\si{\angstrom} (HeII$\lambda$1640) has been indicated as a potential tracer of Pop\ IIIs \citep{Tumlinson_Shull_2000, Tumlinson_2001, Bromm_2001_spectra, Schaerer_2002, Schaerer_2003, Raiter_2010}. Due to their pristine chemical composition, Pop\ III stars are expected in fact to be predominantly massive \citep{Abel_2002, Bromm_2002}, up to $\sim$100s~\si{M_\odot} \citep{Hosokawa_2011, Hirano_2014, Stacy_2016, Chon_2024}, or even $\sim$1000s~\si{M_\odot} \citep{Hirano_2015_UVrad, Hirano_2015_PPS, Susa_2014, Hosokawa_2016, Sugimura_2020, Latif_2022}. This massive component should power a very hard radiation (> 54.4 eV), able to doubly ionise He in the nearby gas, and therefore trigger the HeII$\lambda$1640 line emission through the cascading recombination of HeIII.

While Pop\ III stars are predicted to start forming at Cosmic Dawn, around $z \sim 20 - 30$ \citep{Bromm_2013, Klessen_Glover_2023}, cosmological simulations \citep{Xu_2016_latePopIII, Jaacks_2019, Liu_Bromm_2020_2, Sarmento_2018, Sarmento_Scannapieco_2022, Skinner_Wise_2020, Venditti_2023} and semi-analytical models \citep{Visbal_2020, Trinca_2024} suggest that pristine gas reservoirs hosting Pop\ IIIs might persist down to the Epoch of Reionization (EoR, $z \sim 6 - 10$). There are already a few candidates with tentative HeII detection, possibly indicative of Pop\ IIIs at these epochs \citep{Wang_2024, Maiolino_2024, Vanzella_2023}; however, their confirmation is still pending. In fact, similar candidates (e.g. CR7 at $z \simeq 6.6$, \citealt{Sobral_2015}) have been rejected in the past on both observational \citep{Bowler_2017, Matthee_2017} and theoretical \citep{Pallottini_2015, Agarwal_2016} grounds, underscoring the importance of using combined diagnostics of spectral hardness and low metallicity to confirm the Pop\ III nature of these systems \citep[e.g.][]{Inoue_2011, Zackrisson_2011, Mas-Ribas_2016, Nakajima_Maiolino_2022, Trussler_2023, Katz_2023, Cleri_2023}.

Despite the plethora of models predicting Pop\ III star formation at later cosmic times, many observational challenges may account for the lack of clear detections. In fact, Pop\ III clusters are expected to have low masses \citep[e.g.][]{Bromm_2013} and hence be intrinsically faint, so that it might be difficult to detect them even in extremely magnified systems \citep{Zackrisson_2012, Zackrisson_2015}; their signal may also be further absorbed by inter-stellar dust \citep{Venditti_2023, Curtis-Lake_2023, Roberts-Borsani_2023}.
A significant number of Pop\ III systems at these redshifts might fall outside the field-of-view (FoV) of our instruments, if they reside at the periphery of their hosting dark-matter haloes \citep{Venditti_2023}. 
Finally, the HeII recombination signature is expected to be short-lived, due to the brief lifetime of the most massive stars \citep[$\sim$~few Myr, ][]{Schaerer_2002, Schaerer_2003, Katz_2023}.
Understanding all these challenges is crucial to design robust strategies for a systematic search of Pop\ IIIs during the EoR, with the goal of expanding our pool of available candidates. 

This letter aims to explore all these aspects, by combining the statistics of late Pop\ III clumps inferred from the cosmological simulations introduced in \citet{DiCesare_2023} and \citet{Venditti_2023} with an analytical modelling of the HeII emission arising from Pop\ III stars \citep{Schaerer_2002}. In Section~\ref{sec:method}, we describe our cosmological simulations (Section~\ref{sec:method_simulations}) and the adopted procedure to estimate the HeII luminosity from Pop\ III stellar populations (Section~\ref{sec:method_HeIILuminosity}). In Section~\ref{sec:results} we present our results, i.e.: (i) our predictions of the HeII luminosity, compared with the sensitivity of JWST/NIRSpec in different configurations (Section~\ref{sec:results_sensitivity}); (ii) the expected bias due to the limited FoV (Section~\ref{sec:results_FoV}); (iii) the expected number of HeII-emitting Pop\ III systems in existing JWST surveys (Section~\ref{sec:results_probability}). In Section~\ref{sec:discussion} we critically discuss our findings, with particular reference to the effect of dust absorption/scattering (Section~\ref{sec:discussion_dust}) and to the potential of using HeII for the identification of Pop\ III stars compared to other indicators, e.g., pair-instability supernovae (PISNe, Section~\ref{sec:discussion_PISNe}). Finally, Section~\ref{sec:conclusions} presents our conclusions.

\section{Methodology} 
\label{sec:method}

\subsection{Simulating the cosmological environment}
\label{sec:method_simulations}

The cosmological simulations employed in the present work have been performed with the hydrodynamical code \texttt{dustyGadget} \citep{Graziani_2020}, and they are described in \citet{DiCesare_2023}. They consist of eight simulated volumes (U6 - U13\footnote{Data from U9 and U11 are not included in the present work as these simulations present different snapshot dumpings with respect to the others; in fact, these cubes are less star-forming and hence of lower interest for the present study.}), with a comoving side of $50h^{-1} ~ \si{cMpc}$, a total number of $2 \times 672^3$ particles and a mass resolution for dark matter/gas particles of $3.53 \times 10^7 h^{-1} ~ \si{M_\odot}$/$5.56 \times 10^6 h^{-1} ~ \si{M_\odot}$ each, evolved from $z \simeq 100$ down to $z \simeq 4$. A $\mathrm{\Lambda}$CDM cosmology consistent with \citet{Planck_2015} is assumed ($\Omega_{\mathrm{m,0}} = 0.3089$, $\Omega_{\mathrm{b,0}} = 0.0486$, $\Omega_{\mathrm{\Lambda},0} = 0.6911$, $h = 0.6774$).

Detailed information on the \texttt{dustyGadget} code, and particularly on its innovative self-consistent modelling for dust production and evolution, can be found in \citet{Graziani_2020}. The code extends the original implementation of the SPH code \texttt{Gadget-2} \citep{Springel_2005_Gadget-2}, on top of the improvements to the chemical evolution module from \citet{Tornatore_2007_chemicalFeedback, Tornatore_2007_PopIII}, to molecular chemistry and cooling from \citet{Maio_2007}, and to their coupling with Pop\ III/II formation from \citet{Maio_2010, Maio_2011}.
We here briefly summarize the main features of the adopted prescriptions for star formation and feedback, of particular interest for the present work. 

A two-phase ISM model is implemented for each SPH gas particle, following \citet{Springel_Hernquist_2003}. Stellar particles with a mass of $\sim 2 \times 10^6 ~ \si{M_\odot}$ are generated from gas particles with a number density ($n$) above the threshold $n_\mathrm{th} \simeq 300 ~ \si{cm^{-3}}$; the cold gas phase is depleted into stars at a rate $n_\mathrm{cold}/t_\star$, with $n_\mathrm{cold}$ the cold-phase number density and $t_\star = 2.1 ~ \si{Gyr} \times (n / n_\mathrm{th})^{-1/2}$ the characteristic time-scale of the process. The stellar particles represent stellar populations born in an instantaneous burst with an assigned initial mass function (IMF). Depending on the gas metallicity, below or above a critical metallicity\footnote{Here we assume $Z_\mathrm{\odot} = 0.02$ \citep{Anders_Grevesse_1989}.} ($Z_\mathrm{crit} = 10^{-4} ~ \si{Z_\odot}$, \citealt{Bromm_2001, Maio_2010, Graziani_2020}) we define a stellar population to be Pop\ III or Pop\ II/I, respectively. We assume a Salpeter-shaped IMF \citep{Salpeter_1955} with a mass range of [0.1, 100] ([100, 500])~\si{M_\odot} for Pop~II/I (Pop~III); this results in an average lifetime for Pop\ III stars of $\simeq 3$~Myr (see equation~1 of \citealt{Venditti_2023}). The impact of the contribution of low-mass Pop\ III stars, ($\sim 1 - 40 ~ \si{M_\odot}$, the mass range inferred from stellar archaeology, \citealt{2Iwamoto_2005, Keller_2014, Ishigaki_2014, deBennassuti_2014, deBennassuti_2017, Hartwig_2015, Fraser_2017, Rossi_2021, Magg_2022, Aguado_2023_mostMetalPoorStar}) is extensively discussed in \citet{Venditti_2023, Venditti_2024}. In fact, although the aforementioned studies show that a precise modelling of the low-mass end of the IMF is required to reconstruct the detailed nucleosynthetic pattern of old, metal-poor stars, here we are mostly interested in the high-mass tail that is mainly responsible for He ionisation, because of its hard UV photon budget. However, it is important to emphasize that changing the IMF in a way that influences the power at high masses -- either via changes in its shape or mass range -- can affect our results (see the discussion in Section~\ref{sec:method_HeIILuminosity}).

The gas chemical evolution model is adopted from \citet{Tornatore_2007_chemicalFeedback, Maio_2010, Maio_2011}. We include mass-dependent yields from Pop\ III stars in the range [140, 260] \si{M_\odot}, ending their life as PISNe \citep{Heger_Woosley_2002}, and mass and metallicity-dependent yields from Pop\ II/I stars with low-intermediate mass (long-lived, \citealt{vanDenHoek_Groenewegen_1997}) and high mass (> 8 \si{M_\odot}, dying as core-collapse supernovae, \citealt{Woosley_Weaver_1995}), also considering type Ia supernovae \citep{Thielemann_2003}. For simplicity, we assume that Pop\ II/I stars more massive than 40~\si{M_\odot} and Pop\ III stars outside the PISN range directly collapse into black holes and do not participate in the metal enrichment process. This clearly is an idealization, and effects such as rapid rotation could contribute to enrichment across a broader range of stellar masses \citep[e.g.,][]{Liu_2021}. Dust and metals are spread in the inter-stellar medium (ISM) through a spline kernel. Galactic winds are also modelled following \citet{Springel_Hernquist_2003}, with a constant velocity of $500 ~ \si{km.s^{-1}}$ \citep{Tornatore_2010, Maio_2011}.

The simulations have demonstrated good agreement with available model predictions and observations of the cosmic star-formation-rate/stellar-mass density evolution and with important scaling relations (i.e., the main sequence of star-forming galaxies, the stellar-to-halo mass relation and the dust-to-stellar mass relation), including early JWST data \citep{DiCesare_2023}; we emphasize here that our model is not calibrated on any particular observational set or survey. The simulations have also been employed to investigate C envelopes around merging galaxies as a possible origin of the [CII]158~$\mu$m emission in the circum-galactic medium surrounding individual, resolved galaxies, observed by the ALPINE\footnote{ALMA Large Program to Investigate [CII] at Early Times Survey \citep[\url{http://alpine.ipac.caltech.edu/}]{LeFevre_2020, Faisst_2020, Bethermin_2020}.} survey at $z \sim 4.5$ \citep{DiCesare_2024}. Most notably, they are the largest simulated volumes currently available that include a model for Pop\ III stars, making them a powerful tool to understand the statistics of Pop\ III star formation across cosmic time \citep{Venditti_2023, Venditti_2024}. However, we emphasize that the limited mass resolution, together with the lack of a proper treatment of radiative feedback\footnote{The simulations only include a homogeneous UV background as in \citet{Haardt_Madau_1996} at $z < 6$, hence neglecting the effect of radiative feedback on cosmic star formation at higher redshifts. See appendix~A of \citet{Venditti_2023} for a discussion of the impact of neglecting UV and LW feedback on the overall Pop\ III star formation history, in the considered redshift range and at the considered resolution.}, allows us to provide reliable results only for haloes with a stellar mass\footnote{Corresponding to a number of stellar particles $\gtrsim 20$.} of $\mathrm{log} (M_\star/ \si{M_\odot}) \gtrsim 7.5$ at $6 \lesssim z \lesssim 10$; note that all Pop\ III stars in this mass regime at the considered redshifts are found to be coexisting with Pop\ II stellar components in our simulations \citep{Venditti_2023}. We also currently do not include a model for metal mixing and turbulent metal diffusion below our gas mass resolution \citep[as e.g. in][]{Sarmento_2016, Sarmento_2017, Sarmento_2018, Sarmento_Scannapieco_2022}. We refer the reader to \citet{Venditti_2023, Venditti_2024} for a thorough discussion of these limitations for Pop\ III studies.

\subsection{Computing the HeII luminosity of Pop\ III clumps}
\label{sec:method_HeIILuminosity}

The intrinsic luminosity of the HeII$\lambda$1640 line ($L_\mathrm{HeII}$) emitted from a {Pop\ III clump (i.e. a Pop\ III stellar cluster, represented by a stellar particle in our simulations as defined in Section~\ref{sec:method_simulations}) can be inferred from the mass of the clump ($M_\mathrm{III}$) as follows: 

\begin{equation}
    L_\mathrm{HeII} = \overline{\varepsilon}_\mathrm{HeII} E_\mathrm{HeII} \times M_\mathrm{III},
    \label{eq:HeIILuminosity}
\end{equation}

with $E_\mathrm{HeII} \simeq 1.21 \times 10^{-11}$~erg the energy of a HeII$\lambda$1640 photon and $\overline{\varepsilon}_\mathrm{HeII}$ the average HeII photon emissivity per unit stellar mass of a Pop\ III stellar population. We do not take into account dust attenuation, whose expected impact will be further discussed in Section~\ref{sec:discussion_dust}.

\begin{figure}
    \centering    
    \includegraphics[width=\linewidth]{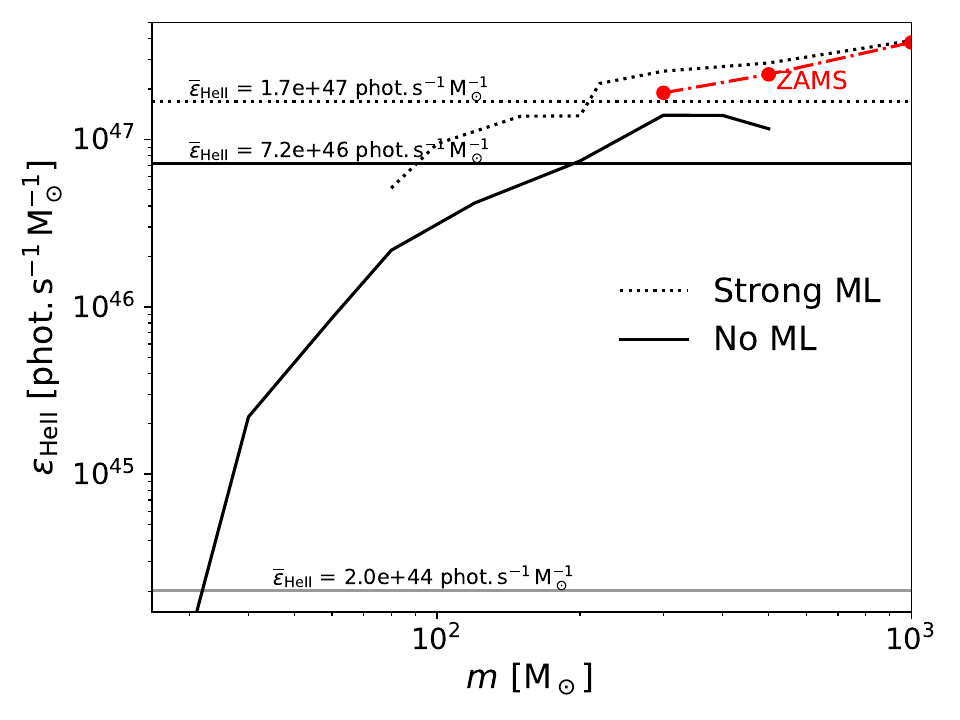}
    \caption{Time-averaged HeII photon emissivity per unit stellar mass $\varepsilon_\mathrm{HeII}$ as a function of the initial Pop\ III mass $m$, from \citet{Schaerer_2002}. The \textit{black, solid line} refers to the model assuming no mass loss (no ML, see table~4 of the original paper), while the \textit{black, dotted line} refers to the model assuming strong mass loss (strong ML, see their table~5). The average emissivity $\overline{\varepsilon}_\mathrm{HeII}$ of Pop\ III stellar populations for the two models with our assumed Salpeter-like IMF in the range [100, 500]~\si{M_\odot} -- i.e., the values adopted in Equation~\ref{eq:avHeIIEmissivity} for the present work -- are indicated on top of the \textit{horizontal, solid/dotted, black lines}; the value for a Salpeter-like IMF in the range [1, 1000]~\si{M_\odot} with no ML is also indicated on top of \textit{horizontal, solid, gray line} (see text for details). The \textit{red, dashed-dotted line} further shows the cases of 300~\si{M_\odot}, 500~\si{M_\odot} and 1000~\si{M_\odot} stars evolving on the ZAMS along the entire stellar lifetime as a reference \citep{Bromm_2001_spectra}, to exemplify how mass loss keeps stellar evolution closer to the ZAMS, enhancing the time-averaged photon emissivity (as in the \textit{black, dotted line}).}
    \label{fig:HeIIEmissivity}
\end{figure}

Figure~\ref{fig:HeIIEmissivity} shows the time-averaged photon production rate $\varepsilon_\mathrm{HeII}$ of individual Pop\ III stars of various masses $m$ (averaged over the lifetime of each star), assuming either strong mass loss (ML) arising from high-mass stars (dotted line) or no mass loss at all (solid line), from tables 5 and 4 of \citet{Schaerer_2002} respectively. We compute $\overline{\varepsilon}_\mathrm{HeII}$ by integrating over the IMF $\phi(m)$: 
\begin{equation}
    \overline{\varepsilon}_\mathrm{HeII} = \frac{\int_{m_\mathrm{low}}^{m_\mathrm{up}} \varepsilon_\mathrm{HeII}(m) \phi(m) \dd{m}}{\int_{m_\mathrm{low}}^{m_\mathrm{up}} \phi(m) \dd{m}}.
    \label{eq:avHeIIEmissivity}
\end{equation}
For our assumed Salpeter-like IMF (whose lower and upper limits are $m_\mathrm{low} = 100 ~ \si{M_\odot}$ and $m_\mathrm{up} = 500 ~ \si{M_\odot}$, Section~\ref{sec:method_simulations}), this results in $\overline{\varepsilon}_\mathrm{HeII} \simeq 1.53 \times 10^{47} / 6.48 \times 10^{46} ~ \si{{phot.}.s^{-1}.M_\odot^{-1}}$ for the case with strong/no mass loss respectively. By considering a wide range of possible IMFs (as in table~1 of \citealt{Venditti_2024}), we find that this value can become up to $\sim 350$ times lower depending on the adopted IMF (particularly, this value is found for a Salpeter-like IMF in the range [1,1000]~\si{M_\odot}, with no mass loss\footnote{As \citet{Schaerer_2002} does not provide results for the HeII emissivity below a mass of 8/5~\si{M_\odot} (for the strong/no-mass-loss case respectively), we conservatively assume that the time-averaged emissivity in Equation~\ref{eq:avHeIIEmissivity} is zero below the available mass range. This is a good approximation for the no-mass-loss case as the HeII emissivity has dropped by more than twelve orders of magnitude between the case of a 500~\si{M_\odot} and 5~\si{M_\odot} stars (see table~4 of \citealt{Schaerer_2002}). We further assume that the HeII emissivity for a 1000~\si{M_\odot} star in the no-mass-loss case - also not provided in \citet{Schaerer_2002} - is the same with respect to a 500~\si{M_\odot} star; in fact, the production rates per unit mass of stars $> 300 ~ \si{M_\odot}$ are found to be essentially independent of stellar mass, within a factor 2 \citep{Bromm_2001_spectra}.})
. Note that mass loss causes the stars to evolve close to the Zero-Age-Main-Sequence (ZAMS) for longer times, resulting in higher time-averaged photon emissivities\footnote{We emphasize that the emission model for Pop\ III stars is not consistent with the feedback model of the simulations. Similarly to \citet{Venditti_2024}, in fact, we only explore how our assumptions on Pop\ IIIs affect the HeII emissivity in post-processing, while a complete discussion would require a self-consistent treatment, also taking into account their impact on the overall star-formation history.}. In fact, the model\footnote{Shown for the cases of 300~\si{M_\odot}, 500~\si{M_\odot} and 1000~\si{M_\odot} Pop\ III stars. See table~1 of \citet{Bromm_2001_spectra} for the case of a 1000~\si{M_\odot} Pop\ III star; the 300~\si{M_\odot} and 500~\si{M_\odot} values are obtained from private communication.} of \citet{Bromm_2001_spectra} -- assuming Pop\ III stars always evolving along the ZAMS -- lies closer to the strong-mass-loss case (red, dashed-dotted line).

As in \citet{Venditti_2024}, we consider the possibility of a Pop\ III mass $M_\mathrm{III}$ in our Pop\ III clumps (in Equation~\ref{eq:HeIILuminosity}) that is lower than our resolution element $M_\mathrm{III,res} \sim 2 \times 10^6 ~ \si{M_\odot}$ (i.e. the mass of a Pop\ III stellar particle in our simulations), by introducing an efficiency factor $\eta_\mathrm{III} < 1$:
\begin{equation}
    M_\mathrm{III} = \eta_\mathrm{III} M_\mathrm{III,res}.
    \label{eq:popIIIResolutionElement}
\end{equation}
By interpreting $M_\mathrm{III,res}$ as the amount of extremely metal-poor gas above our density threshold that is available for star formation, and $M_\mathrm{III}$ as the amount of stellar mass actually produced in a single star-formation event\footnote{We note that, although in the present study we focus on Pop\ III stellar clusters, the formation of isolated Pop III stars with masses between $\sim 100$ and 500~\si{M_\odot} has also been considered in the literature \citep[e.g. by][]{Katz_2023}, and previous studies \citep[e.g.][]{Rydberg_2013, Windhorst_2018} seem to indicate that such individual Pop III stars would be too faint to be observed unless subject to extreme gravitational lensing.}, we can place a lower limit on $\eta_\mathrm{III} \sim 0.01$ from simulations describing Pop\ III star formation in the first mini-haloes (see e.g. \citealt{Bromm_2013} and references therein). However, higher values might be found in more massive haloes at later times, which are expected to host more efficient star-formation sites. In fact, a past simulation \citep{Greif_2008} of primordial gas collapsing into an atomic-cooling halo at $z \sim 10$ shows that the gas experiences a boost in ionization (e.g. through shocks) resulting in more efficient cooling through the HD channel \citep{Bromm_2009}; this intermediate regime -- previously referred to as Pop\ III.2 or Pop\ II.5 -- between the very first episodes of star formation and later Pop\ II/I star formation has been predicted to yield higher star-formation efficiencies, even a factor $\sim$10 higher than star formation in mini-haloes at Cosmic Dawn \citep{Greif_Bromm_2006}. In the absence of tight constraints in the mass regime we are currently probing, we explore values up to $\eta_\mathrm{III} \sim 0.1$. We note that a similar value would also be implied considering the ratio of the mass\footnote{The mass estimate from \citet{Maiolino_2024} has been updated with respect to \citet{Venditti_2024} to match the accepted version of the paper.} inferred for the supposed Pop\ III clusters in GN-z11 at $z = 10.6$ \citep[$\sim 2-2.5 \times 10^5 ~ \si{M_\odot}$, ][]{Maiolino_2024} and RXJ2129-z8HeII at $z \simeq 8.2$ \citep[$7.8 \pm 1.4 \times 10^5 ~ \si{M_\odot}$, ][]{Wang_2024} with respect to our resolution element.

Using the HeII line alone to distinguish between the underlying emission model for Pop\ III stars is challenging, due to the degeneracy among all the uncertain parameters that determine the HeII luminosity. However, in Section~\ref{sec:results_sensitivity} we provide a broad range of possible values for $L_\mathrm{HeII}$ to offer clues on the sensitivity required to rule out the presence of Pop\ III stars in high-$z$ galaxies, taking into account the considerable uncertainties on their nature.

\section{Results} 
\label{sec:results}

\subsection{HeII luminosity vs. JWST/NIRSpec sensitivity}
\label{sec:results_sensitivity}

\begin{figure*}
    \centering
    \includegraphics[width=\linewidth]{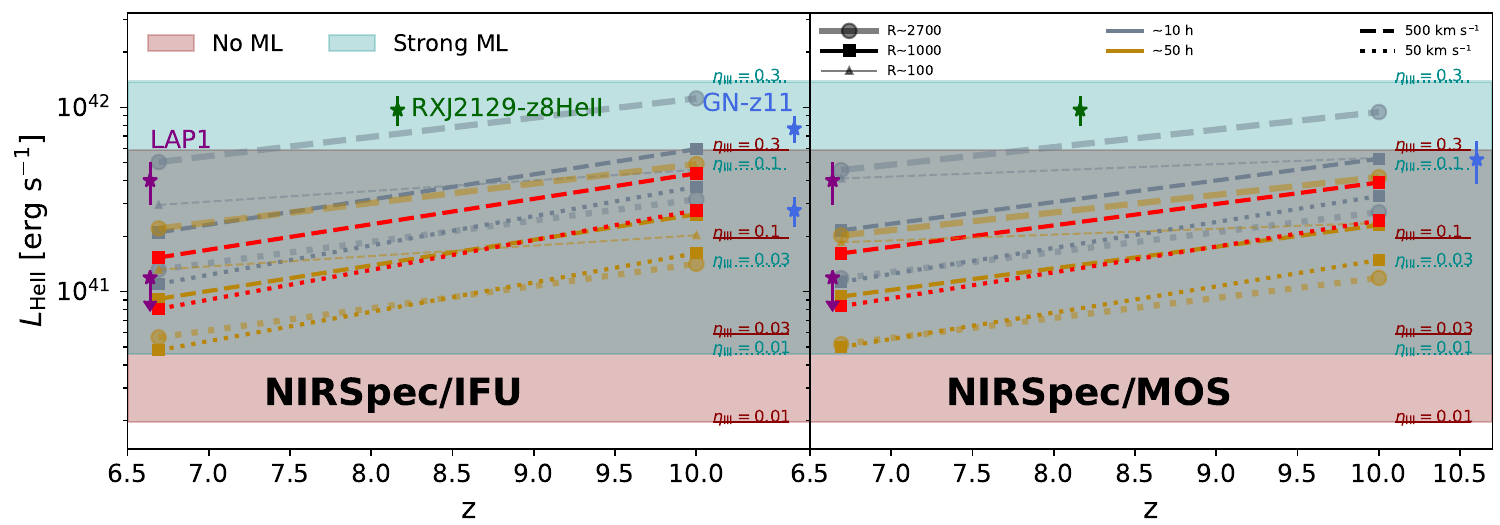}
    \caption{Average HeII line luminosity ($L_\mathrm{HeII}$, \textit{shaded regions)} arising from Pop\ III-hosting \texttt{dustyGadget} galaxies at $z \sim 6.7 - 10$, assuming a Salpeter-like IMF in the range [100, 500]~\si{M_\odot}, compared with available observational candidates, i.e., LAP1 at $z \simeq 6.6$ \citep[\textit{purple stars}]{Vanzella_2023}, RXJ2129-z8HeII at $z \simeq 8.2$ \citep[\textit{green star}]{Wang_2024}, and GN-z11 at $z \simeq 10.6$ \citep[\textit{blue stars}]{Maiolino_2024}. The \textit{dark-red/dark-cyan shaded regions} refer to the models assuming no/strong mass loss (ML) respectively, with variable Pop\ III formation efficiency $\eta_\mathrm{III}$ from 0.01 (\textit{lower end} of the regions) to 0.3 (\textit{upper end}), and considering the average Pop\ III mass in \texttt{dustyGadget} galaxies at these redshifts in Equation~\ref{eq:HeIILuminosity}. The \textit{oblique lines} show the sensitivity of JWST/NIRSpec corresponding to an integrated $\mathrm{S/N} \sim 3$ line detection in both the IFU (\textbf{left}) and MOS (\textbf{right}) observing modes at $z \simeq 6.7$ and $z \simeq 10$ for different configurations, i.e., different resolving powers ($R \sim 1000/2700/100$, with different \textit{thickness} and \textit{markers}), exposure times ($\sim 10/50$~h, \textit{grey/golden}) and line widths (500/50~\si{km.s^{-1}}, \textit{dashed/dotted}). The corresponding lines for an integrated $\mathrm{S/N} \sim 5$ line detection with medium resolving power and a $\sim$~50~h exposure are also shown in \textit{red}, with the same \textit{linestyles} as the $\mathrm{S/N} \sim 3$ case for the two assumed line widths.}
    \label{fig:HeIILuminosityVsRedshift}
\end{figure*}

Figure~\ref{fig:HeIILuminosityVsRedshift} provides predictions of the HeII luminosity ($L_\mathrm{HeII}$) arising from \texttt{dustyGadget} galaxies hosting Pop\ IIIs as a function of redshift $z$, considering different assumptions on the Pop\ III formation efficiency $\eta_\mathrm{III}$ and mass loss. We estimate the average Pop\ III mass ($M_\mathrm{III} \simeq 2.24 \times 10^6 ~ \si{M_\odot}$) in Pop\ III-hosting galaxies above $M_\star \simeq 10^{7.5} ~ \si{M_\odot}$ between redshifts $\sim 6$ and $\sim 10$, and compute the resulting $L_\mathrm{HeII}$ from Equation~\ref{eq:HeIILuminosity}. The upper end of the shaded regions in the plots is associated with the models assuming strong mass loss and high $\eta_\mathrm{III}$ (up to a value\footnote{This upper limit is derived by considering the ratio of the Pop\ III mass inferred for RXJ2129-z8HeII ($\sim 7.8 \times 10^5 ~ \si{M_\odot}$) to our typical Pop\ III stellar particle masses.} $\eta_\mathrm{III} = 0.3$), while the lower end is associated with no mass loss and low $\eta_\mathrm{III}$ (down to $\eta_\mathrm{III} = 0.01$). The $L_\mathrm{HeII}$ spread is dominated by the uncertainty on $\eta_\mathrm{III}$, while the presence/absence of mass loss only accounts for a factor of order unity. As discussed in Section~\ref{sec:method_HeIILuminosity}, considering a different IMF can lead to lower luminosities, up to a factor $\sim 1 / 350$; moreover, these results do not account for dust absorption, which might lead to very high attenuations (up to a factor $\sim 10^{-9}$) along particularly unfavourable lines-of-sight, although along more typical lines-of-sight less than 10\% of the flux would be absorbed (see the discussion in Section~\ref{sec:discussion_dust}). The purple, green and blue stars provide a comparison with the luminosity of available observational candidates, i.e., LAP1\footnote{Note that, while the authors provide constraints on the expected HeII flux in LAP1, the identification of the observed feature as a HeII$\lambda$1640 line is weakened by the presence of a small blueshift relative to the Balmer lines. Moreover, the measured flux would require a quite extreme Pop\ III scenario. 
Hence, the authors conservatively consider the line non-detected and derive an upper limit on the HeII flux, also shown in the plot.} at $z \simeq 6.6$ \citep{Vanzella_2023}, RXJ2129-z8HeII at $z \simeq 8.2$ \citep{Wang_2024} and GN-z11\footnote{\label{foot:GN-z11}The two points for GN-z11 in the left panel indicate two different NIRSpec/IFU measures, considering a small aperture around the HeII clump (bottom point), and a larger aperture aimed at also capturing an additional, more extended emission, possibly coming from a fainter, less significant clump (top point). The point in the right panel refers instead to the NIRSpec/MOS measure. Although the two measures are consistent in terms of wavelength, comparing the fluxes is non-trivial due to the uncertainty on the exact location of the MSA shutter, and hence on the covered fraction of the putative HeII clump \citep{Maiolino_2024}.} at $z \simeq 10.6$ \citep{Maiolino_2024}. 

The oblique lines indicate sensitivity limits for JWST/NIRSpec \citep{Jakobsen_2022} in both the Integral Field Unit (IFU) and Multi-Object Spectroscopy (MOS) modes at $z \simeq 6.7$ and $z \simeq 10$ for different configurations. 
The limits are computed using version 4.0 of the JWST Exposure Time Calculator\footnote{\url{https://jwst.etc.stsci.edu}}, assuming a point source with no continuum\footnote{If the continuum is detected, this will effectively boost the line. For example, by considering a flat continuum of $\sim 1 - 10$~nJy, the total S/N at the wavelength of the line is approximately enhanced by a quantity of the order of the S/N for the detection of the continuum itself.} and a line centered at $\lambda \simeq 1.64 \times [(1 + z)/10] ~ \mu\si{m}$, plus medium background\footnote{Backgrounds in the ETC are obtained using a background model generator accounting for the various components that contribute to the JWST background \citep{Rigby_2023}. Particularly, a medium background accounts for the 50\% percentile over the period of visibility in a given celestial position. We used as a reference the position of the HeII clump in GN-z11, for consistency among the various calculations. By considering for example the position of LAP1 and RXJ2129-z8HeII, we find a variation up to $\sim 8\%$ (specifically, for the position of RXJ2129-z8HeII) in the expected S/N at the same limiting flux.}. We adopt an operational integrated signal-to-noise (S/N) threshold of $\sim 3$ to determine the minimum observable line flux, which depends on the specific observational setup as well as the chosen line width. We explore two possible values for the line width, $\Delta v = 500/50 ~ \si{km.s^{-1}}$ (dashed/dotted lines), corresponding respectively to typical virial velocities in high-$z$ galaxies ($\Delta v \simeq 50 ~ \si{km.s^{-1}}$), and to a more extreme scenario\footnote{A similar value of $\Delta v \simeq 428 ~ \si{km.s^{-1}}$ has been found for the HeII$\lambda$1640 line detected in the lensed galaxy RXCJ2248-ID at $z = 6.1$ \citep{Topping_2024}, whose spectrum broadly resembles that of GN-z11 \citep{Maiolino_2024}, minus the AGN signatures.}, typical of feedback-generated velocities, such as supernova-driven outflows ($\Delta v \simeq 500 ~ \si{km.s^{-1}}$).
We further consider observations with two exposure times, $t \simeq 10/50$~h (grey/golden lines), with the appropriate grating/filter pair depending on the redshifted wavelength of the line at medium/high resolution (i.e., resolving power $R = 1000/2700$) and with the Prism/CLEAR at low resolution\footnote{For the PRISM/CLEAR, only the 500~\si{km.s^{-1}} case is shown, as the instrument cannot discriminate emission lines with $\Delta v \lesssim 5000 ~ \si{km.s^{-1}}$ at these wavelengths, and hence using a lower $\Delta v$ does not change the results.} ($R = 100$). 

In the IFU mode, the instrument is centered on the source with an aperture\footnote{As we aim for the smallest possible aperture in order to optimise the sampled flux at the center, we consider a value of the order of the Point Spread Function (PSF) Full Width Half Maximum (FWHM) at $\lambda \sim 1.5 ~ \mu\si{m}$ (as a reference, note that the line is redshifted at $\lambda \simeq 1.3/1.8 ~ \mu\si{m}$ at $z \simeq 6.7/10$ respectively).} of 0.09'', while the sky annulus\footnote{The choice of the sky annulus used for background subtraction does not change our results appreciably, as we are considering a uniform background.} spans a range between 0.3'' and 0.9''. In the MOS mode, we select a three-shutters (-1,0,1) slitlet shape, with the source placed in shutter 0 and the Micro-Shutter Assembly (MSA) located in quadrant 3 center; we apply the MSA full shutter extraction strategy for background subtraction. The Improved Reference Sampling and Subtraction (IRS$^2$, \citealt{Rauscher_2012}) readout pattern is employed for both cases.

It is evident that even with $\sim 10$~h observations and considering the narrow-line, best-case scenario at all the available spectral resolutions, we would only be able to capture very luminous Pop\ III systems ($L_\mathrm{HeII} \gtrsim 10^{41} ~ \si{erg.s^{-1}}$, i.e. assuming $\eta_\mathrm{III} \gtrsim 0.02/0.06$ in the strong/no mass loss case). All the observed candidates lie in fact in the upper part of this plot. Very low-luminosity systems ($\lesssim 4 \times 10^{40} ~ \si{erg.s^{-1}}$) would be missed even in the deepest exposures ($\sim 50$~h). However, it is to be noted that more favourable conditions are possible. For example, while here we conservatively assumed a medium background, the very low background during the NIRSpec/IFU observation of GN-z11 allowed the detection of a fainter HeII line than expected (with a $\simeq 10.6$~h exposure, in the small aperture). Moreover, the line appears unresolved in the G235M/F170LP grating/filter pair ($R \sim 1000$), meaning even narrower lines (with higher S/N) may be found.

\subsection{Pop\ IIIs outside JWST/NIRSpec field-of-view}
\label{sec:results_FoV}

Another factor contributing to the potential oversight of Pop\ III systems is their placement outside the FoV of our instruments. As star formation is typically more efficient in the dense, central regions of galaxies \citep[e.g, ][]{Carrasco_2010, vanDokkum_2014}, peripheral areas tend to evolve at a slower pace, preserving their chemically pristine state for extended periods of time; additionally, these regions may experience gas infall from the external environment. Pristine star-forming regions may also reside in small satellites at the periphery of the same dark matter halo. As a result, Pop\ III stars might be found as far as $\sim 20$~kpc from the galactic centre (see figure~10 of \citealt{Venditti_2023}), especially in regions surrounding massive, evolved galaxies, which have undergone prolonged periods of star formation. \citet{Maiolino_2024}, for example, find a potential Pop\ III clump at a distance $\sim$~2~kpc from the host galaxy of GN-z11 ($M_\star \sim 8 \times 10^8 ~ \si{M_\odot}$). 

\begin{figure}
    \centering
    \includegraphics[width=\linewidth]{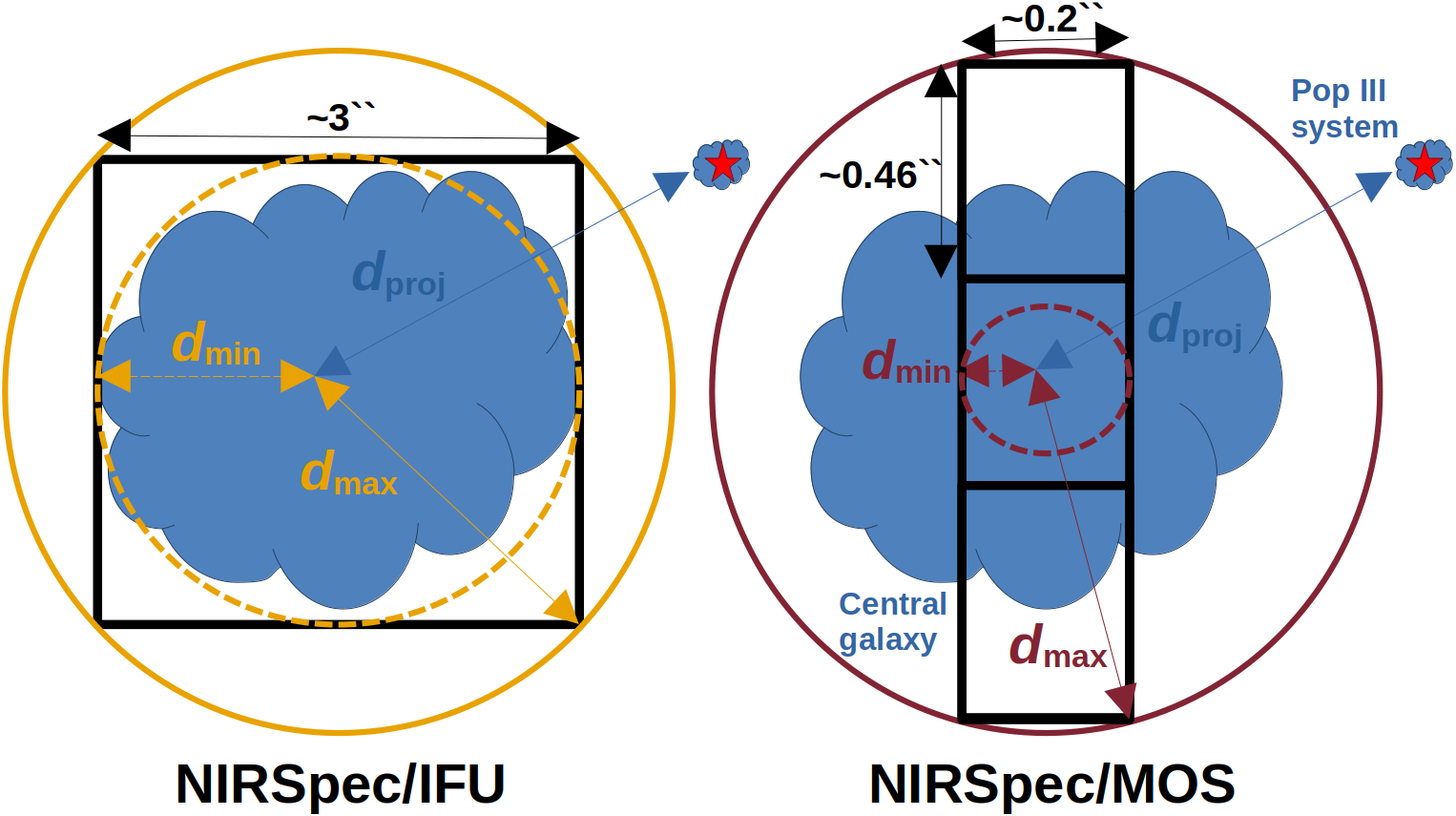}
    \caption{Illustration of a single JWST/NIRSpec pointing centered on Pop\ III-hosting galaxies (\textit{blue clouds}) in the IFU/MOS modes, respectively (\textbf{left/right}; in the MOS mode, three consecutive shutters are considered). The distances $d_\mathrm{proj}$ of the peripheral Pop\ III clusters (\textit{small blue clouds + red stars}) from the galactic center in the displayed projection are shown as \textit{blue arrows}, while the largest distances such that the clusters would certainly fall within the FoV independent of the instrument's orientation ($d_\mathrm{min}$), or that conversely they would fall within for at least one possible orientation ($d_\mathrm{max}$), are shown as \textit{dashed/solid arrows}, spanning the whole \textit{dashed/solid circumferences}. The dimensions of NIRSpec/IFU FoV and of the three shutters of NIRSpec/MOS are also indicated in the figure (not to scale). Depending on the orientation of the galaxies with respect to our instruments, peripheral Pop\ III systems might fall out of the FoV and would thus be missed in observations.}
    \label{fig:FoVScheme}
\end{figure}

We estimate the number of Pop\ III systems that might be missed in a single pointing of JWST/NIRSpec centered on \texttt{dustyGadget} haloes. Note that two degrees of freedom need to be taken into account in this calculation:
\begin{enumerate}
    \item the orientation of the galaxy with respect to our instruments. In fact, when projecting a three-dimensional galaxy onto a two-dimensional image on the sky along an arbitrary direction, the hosted Pop\ III cluster will be observed at a distance $d_\mathrm{proj}$ from the galactic center which is at most equal to the three-dimensional distance $d_\mathrm{3D}$. Particularly, there is always a direction along which the Pop\ III cluster is exactly aligned with the center along our line-of-sight to the source. For each simulated galaxy, we consider the worst possible projection in which $d_\mathrm{proj} = d_\mathrm{3D}$, hence providing an upper limit on the number of Pop\ III systems that we expect to be missing when pointing towards the galactic center. We further consider the case of an average line-of-sight ($\langle d_\mathrm{proj} \rangle = d_\mathrm{3D} \times \pi/4$);
    \item the orientation of the instrument. As shown in Figure~\ref{fig:FoVScheme}, we consider two cases: (i) a Pop\ III system is ``missed'' when we are never able to see it however we rotate the FoV, i.e., when $d_\mathrm{proj} > d_\mathrm{max}$ (falling out of the solid circumferences encompassing all the possible orientations of the instrument); (ii) a Pop\ III system is ``potentially missed'' when it might be missed depending on the particular orientation of the FoV, i.e., $d_\mathrm{proj} > d_\mathrm{min}$ (falling out of the dashed circumference enclosing the area always covered with a random, fixed orientation of the instrument). For the IFU/MOS NIRSpec modes (the latter assumed in a three-shutters configuration), we have $d_\mathrm{max}\simeq 2.1/1.0''$ and $d_\mathrm{min} \simeq 1.5/0.1''$. 
\end{enumerate}

\begin{figure*}
    \centering
    \includegraphics[width=\linewidth]{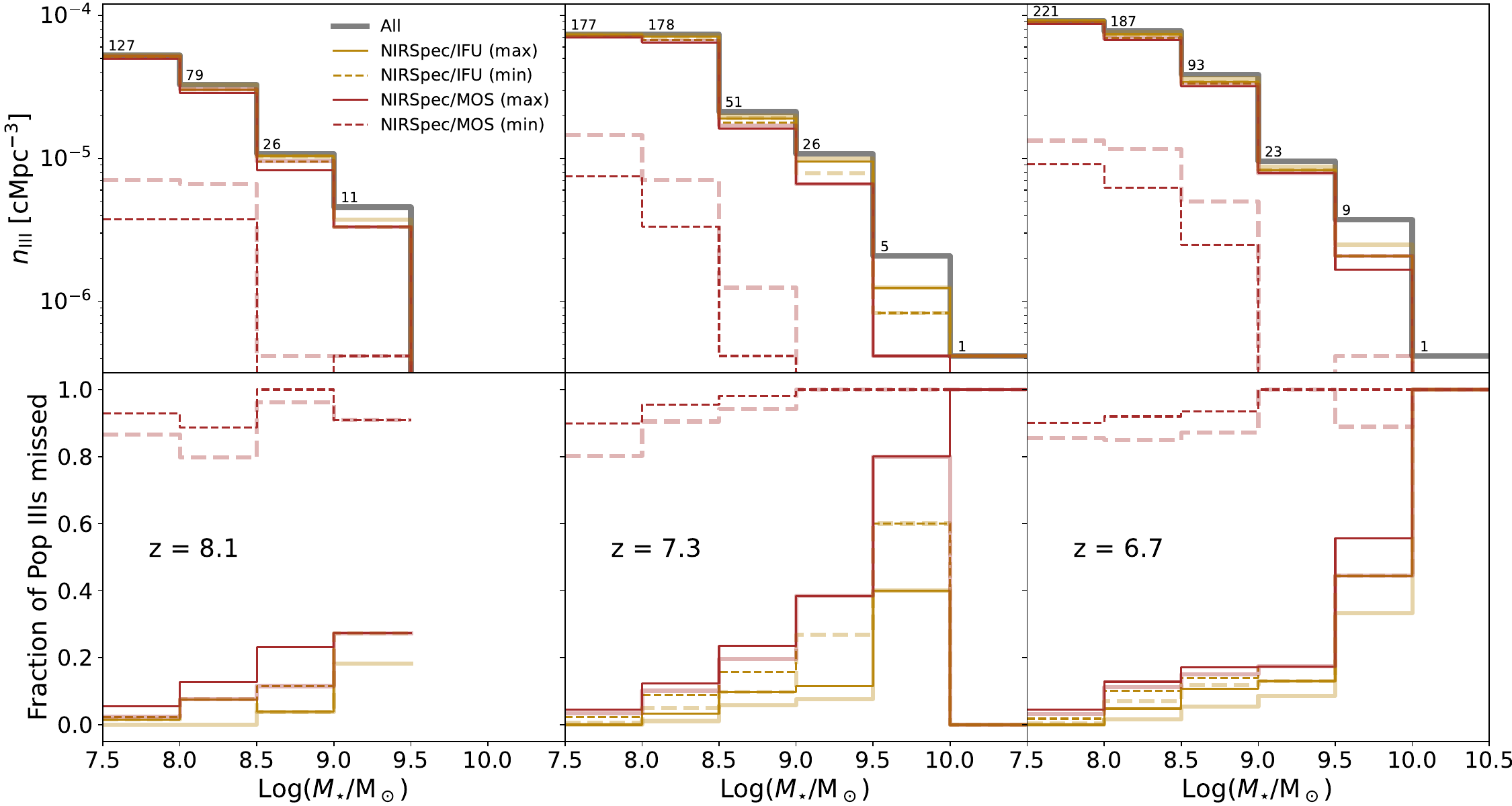}
    \caption{\textbf{Top panels:} number density of Pop\ III particles ($n_\mathrm{III}$) that we expect to find at a given redshift $z$ in haloes within a given range of stellar mass $M_\star$ as a function of $M_\star$, computed in six bins of $M_\star$ (with a spacing of 0.5~dex in the range $7.5 \leq \mathrm{Log} M_\star / \si{M_\odot} < 10.5$). The total $n_\mathrm{III}$ is shown as a \textit{grey, solid line}, while the $n_\mathrm{III}$ found considering a single pointing of JWST/NIRSpec centered on the host galaxy in the IFU/MOS mode, respectively, are shown as \textit{golden/brown lines}. The \textit{solid/dashed} linestyle refer to the best/worst-case scenario for the orientation for the instrument, considering the worst line-of-sight to the source and an average line-of-sight (the latter are shown with \textit{thick, transparent} lines; see text and Figure~\ref{fig:FoVScheme} for details). The total number of Pop\ III particles found in each bin is also indicated on top of the bins. \textbf{Bottom panels}: fraction of Pop\ IIIs missed in JWST/NIRSpec pointings, with same colour/linestyle convention. Results are shown for the combined simulated volumes U6, U7, U8, U10, U12, and U13 at redshifts $z = 8.1$ (\textit{left panels}), $z = 7.3$ (\textit{middle panels}) and $z = 6.7$ (\textit{right panels}). We find that a significant number of Pop\ III systems can be overlooked in these configurations, especially in high-mass galaxies, although this problem appears to be less severe with NIRSpec/IFU thanks to its larger FoV.}
    \label{fig:FoV}
\end{figure*}

The top panels of Figure~\ref{fig:FoV} show the number density of Pop\ III particles that would be found within NIRSpec FoV in all the considered configurations, compared to the total; we remark that the average lifetime of Pop\ III stars with our assumed IMF is $\simeq 3$~Myr (see Section~\ref{sec:method_simulations}), which is consistent with predictions for the lifetime of HeII signatures found by previous works \citep[e.g.][]{Schaerer_2002, Schaerer_2003, Katz_2023}. Results are shown at $z = 8.1$, 7.3 and 6.7 in bins\footnote{The total number of Pop\ III particles found in each bin among all the simulated cubes is indicated in the plots, serving as a cautionary note regarding the limited statistics at the highest-stellar-mass bins. More reliable results would necessitate even larger simulated boxes or more simulated volumes.} of stellar mass\footnote{For reference, the relation between stellar mass and dark matter mass in \texttt{dustyGadget} galaxies is shown in figure~7 of \citet{DiCesare_2023}.} $M_\star$. The bottom panels illustrate, conversely, the fraction of Pop\ IIIs missed in all the aforementioned scenarios. We find that a significant number of Pop\ III systems can be overlooked in these configurations, especially in high-mass galaxies, although this problem appears to be less severe with NIRSpec/IFU thanks to its larger FoV. 

Note that, in the case of GN-z11, the HeII$\lambda$1640 emission found to peak at about $d_\mathrm{proj} = 0.5''$ from the center (i.e. $d_\mathrm{min} < d_\mathrm{proj} < d_\mathrm{max}$ for the MOS mode), would have been missed with a different orientation of the MSA, while it is always included in the FoV of NIRSpec/IFU ($d_\mathrm{proj} < d_\mathrm{min}$). Interestingly, the small size of the MSA slits have also been found to lead to a possible underestimation of the Ly$\alpha$ flux in Ly$\alpha$ emitters in the presence of a spatial offset between the UV and Ly$\alpha$ emission, or of an extended diffuse Ly$\alpha$ emission \citep{Nakane_2024, Napolitano_2024}.

We find a total number density of Pop\ III systems in galaxies above $M_\star = 10^{7.5} ~ \si{M_\odot}$ of $\sim 1.0/1.8/2.2 \times 10^{-4} ~ \si{cMpc^{-3}}$ at $z = 8.1/7.3/6.7$ respectively. Even when neglecting losses due to geometrical effects, this is much lower than the minimum number density predicted e.g. by \citet{Vikaeus_2022} to detect at least one Pop III system in a single, blind NIRSpec survey with an area of $0.0034 ~ \si{deg^2}$ and a sensitivity of $1.3 \times 10^{-19} ~ \si{erg.s^{-1}.cm^{-2}}$: from their figure~3, even assuming a high Pop\ III mass of $\sim 4.4 \times 10^5 ~ \si{M_\odot}$ (orange curve), the required number density is of $\sim 3.5 \times 10^{-2}$/$8.9 \times 10^{-3}$/$1.8 \times 10^{-3} ~ \si{cMpc^{-3}}$ at the same considered redshift points. This might be an indication that blind spectroscopic surveys are not the most efficient strategy to look for Pop\ III stars in massive galaxies, and that more care is needed to select promising candidates/environments\footnote{For example, the same authors find that a single typical cluster lens is about 20 times more effective for a spectroscopic detection of Pop\ IIIs than the considered wide-field surveys. In fact, the smaller survey area ($\sim 0.082 ~ \si{arcmin^2}$) is compensated by higher probabilities to achieve very high magnifications.}, although note that these results are strongly model-dependent\footnote{In the fiducial model of \citet{Vikaeus_2022}, they assume a constant star-formation rate over a time scale of 10~Myr, with stellar populations formed according to a log-normal IMF in the range [1,500]~\si{M_\odot}, with width $\sigma = 1$ and a characteristic mass of 60~\si{M_\odot}. These assumptions result in a much lower HeII luminosity of $\sim 2.64 \times 10^{40} ~ \si{erg.s^{-1}}$ for a Pop\ III mass of $\sim 4.4 \times 10^5 ~ \si{M_\odot}$ with respect to Equation~\ref{eq:HeIILuminosity}, which requires a magnification $\sim 3-4$ to be observable at the considered sensitivity and redshifts.}.

\subsection{Expected Pop\ III systems in JWST surveys}
\label{sec:results_probability}

\begin{figure}
    \centering
    \includegraphics[width=\linewidth]{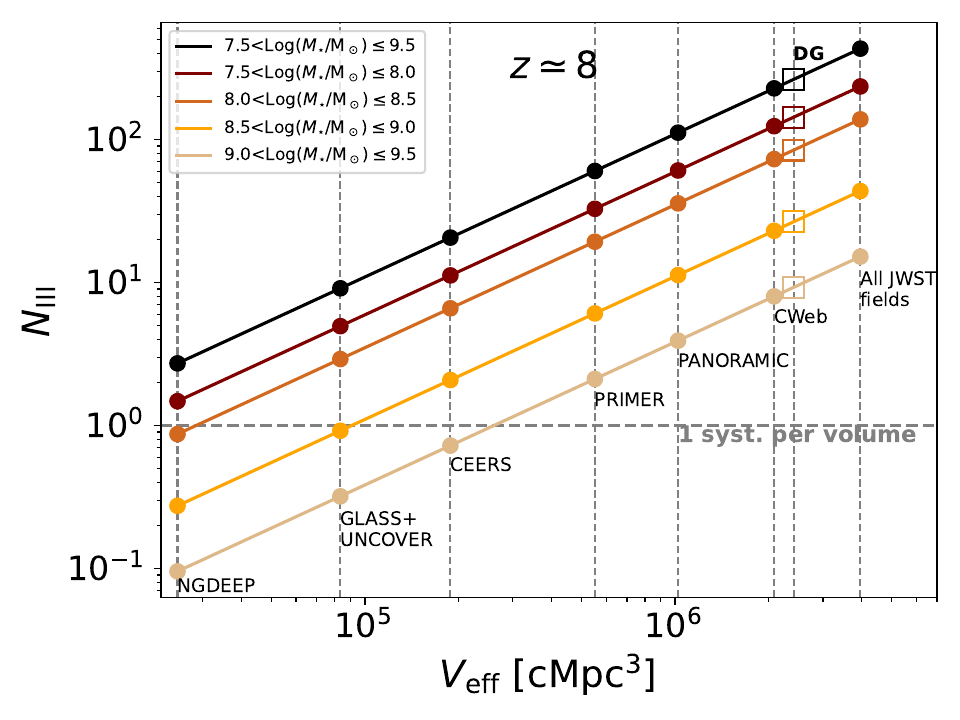}
    \caption{Number of potentially observable HeII-emitting Pop\ III systems $N_\mathrm{HeII}$ as a function of the effective survey volume $V_\mathrm{eff}$ at $z = 8.1$ (see text for details). \textit{Vertical, dashed lines} indicate the effective volume of selected JWST surveys and their cumulative volume at $z \simeq 8$, with $\Delta z = 1$ (\textit{filled circles}, see text for details); the number of Pop\ III systems found in our six \texttt{dustyGadget} cubes is also indicated in the plot (DG, \textit{empty squares}). A \textit{horizontal, dashed line} further indicates the reference value of 1 system per volume. The \textit{black line} refers to the number found in all haloes with $7.5 < \mathrm{Log}(M_\star/\si{M_\odot}) \leq 9.5$, while the \textit{coloured lines} refer to the number found in haloes of different stellar mass bins (see Figure~\ref{fig:FoV}).}
    \label{fig:HeIIStat_vol}
\end{figure}

Figure~\ref{fig:HeIIStat_vol} shows the predicted number $N_\mathrm{III}$ of HeII-emitting Pop\ III systems in galaxies of various masses ($7.5 < \mathrm{Log}(M_\star/\si{M_\odot}) \leq 9.5$) at $z = 8.1$ that are potentially observable with a single pointing of JWST/NIRSpec in its IFU mode, as a function of effective survey volume $V_\mathrm{eff}$. We consider the systems that would fall within the FoV in the worst possible projection\footnote{Although note that considering an average line-of-sight as in Figure~\ref{fig:FoV} barely changes the results.} ($d_\mathrm{proj} = d_\mathrm{3D}$) with any orientation of the instrument ($d_\mathrm{proj} < d_\mathrm{min}$; see Section~\ref{sec:results_FoV}), hence this should be considered as a lower limit. Specifically, we multiply the values of $n_\mathrm{III}$ from the simulations (first panel of Figure~\ref{fig:FoV}, golden, dashed line) by the comoving volume of JWST surveys at $z \simeq 8$, with $\Delta z = 1$: 
\begin{enumerate}
    \item the Next Generation Deep Extragalactic Exploratory Public (NGDEEP) Survey \citep{Finkelstein_2021, Pirzkal_2023, Bagley_2024};
    \item the Grism Lens-Amplified Survey from Space (GLASS\footnote{\url{https://glass.astro.ucla.edu/ers/}}, \citealt{Treu_2017, Treu_2022, Castellano_2022});
    \item the Ultradeep NIRSpec and NIRCam ObserVations before the Epoch of Reionization (UNCOVER\footnote{\url{https://jwst-uncover.github.io/}}, \citealt{Bezanson_2022, Furtak_2023, Weaver_2024});
    \item the Cosmic Evolution Early Release Science Survey (CEERS\footnote{\url{https://ceers.github.io/ceersi-first-images-release}}, \citealt{Finkelstein_2017, Finkelstein_2022, Finkelstein_2023});
    \item the Public Release IMaging for Extragalactic Research (PRIMER\footnote{\url{https://primer-jwst.github.io/}}) survey \citep{Dunlop_2021};
    \item the PANORAMIC survey \citep{Williams_2021};
    \item the Cosmic Evolution Survey (COSMOS-Web\footnote{\url{https://cosmos.astro.caltech.edu/}}, \citealt{Casey_2023}).
\end{enumerate}

Our simulations suggest that more than 400 Pop\ III systems could be discovered in galaxies with $M_\star \gtrsim 10^{7.5} ~ \si{M_\odot}$ around $z \simeq 8$, within all these combined JWST fields, and more than one system is supposed to be found within each individual survey. However, we remark that this number is derived from purely geometrical considerations: it indicates the number of Pop\ III sources - selected from a given survey - that would fall within NIRSpec FoV when pointing towards the centre of their galactic host, while issues related to the intrinsic faintness of the sources and dust absorption/scattering are only broadly discussed in Section~\ref{sec:results_sensitivity} and Section~\ref{sec:discussion_dust}, respectively. Costly spectroscopic follow-up is in fact required for their identification. Moreover, the number of HeII emitters above the sensitivity limits for a given instrumental setup strongly depends on the assumed model for Pop\ III star formation (see Section~\ref{sec:results_sensitivity}). As an example, only $\sim 75/85 \%$ of the whole range of luminosities spanned by the models in Figure~\ref{fig:HeIILuminosityVsRedshift} would be covered with a $\sim 10$~h exposure at medium spectral resolution, assuming a line width of 500/50~\si{km.s^{-1}} respectively, while up to $\sim 90/95 \%$ would be covered with a $\sim 50$~h exposure.

These Pop III models are likely not equiprobable in reality. Depending on the actual nature of Pop\ III stars, fainter Pop\ III models might be favoured, causing a large fraction of these systems to only be accessible through very deep exposures; this might be especially true for the more numerous low-mass galaxies, which might be associated with lower star-formation efficiencies and hence lower HeII luminosities. As discussed in Section~\ref{sec:method_HeIILuminosity}, both the Pop\ III-formation efficiency and IMF could in fact vary depending on environmental conditions, such as the mass of the dark matter host where the Pop\ IIIs are formed. We emphasize the need for more in-depth studies of Pop\ III star formation in mini-haloes versus Ly-$\alpha$ cooling haloes, that would allow us to infer a probability distribution function for the values of the two parameters $\overline{\varepsilon}_\mathrm{HeII}$ (Equation~\ref{eq:avHeIIEmissivity}) and $ \eta_\mathrm{III}$ (Equation~\ref{eq:popIIIResolutionElement}) -- and hence for the HeII luminosity $L_\mathrm{HeII}$ (Equation~\ref{eq:HeIILuminosity}) -- as a function of host mass.

\section{Discussion}
\label{sec:discussion}

\subsection{Dust absorption/scattering}
\label{sec:discussion_dust}

In Figure~\ref{fig:HeIILuminosityVsRedshift} we considered the intrinsic HeII emission arising from Pop\ III star-forming regions. However, light can be removed from our lines-of-sight to the sources because of both absorption and scattering by interstellar dust, which should be accounted for via an additional factor $\sim \exp{-\tau}$ in Equation~\ref{eq:HeIILuminosity}, with $\tau$ the dust optical depth at $\lambda = 1640 ~ \si{\angstrom}$. Models of dust mixtures reproducing the observed extinction in the Milky Way\footnote{\url{https://www.astro.princeton.edu/~draine/dust/dust.html}} \citep{Weingartner_Draine_2001, Li_Draine_2001, Draine_2003_interstellarDust, Draine_2003_dustScatteringUV, Draine_2003_dustScatteringX, Glatzle_2019} predict in fact non-negligible values of the dust absorption cross section per unit dust mass ($\sim 30 \%$ of the peak value\footnote{A maximum dust absorption cross section per unit dust mass of $\simeq 1.5 \times 10^5 ~ \si{cm^2.g^{-1}}$ is found assuming an extinction ratio $R_\mathrm{V} = 3.1$ (see e.g. Figure 1 of \citealt{Glatzle_2019}).})
and of the albedo/scattering asymmetry parameter ($\sim 0.4/0.6$) at $\lambda = 1640 ~ \si{\angstrom}$.

Candidate HeII emitters such as GN-z11 \citep{Jiang_2021, Tacchella_2023}, RXJ2129-z8HeII \citep{Wang_2024}, LAP1 \citep{Vanzella_2023} and RXCJ2248-ID \citep{Topping_2024} are consistent with essentially no dust attenuation. However, recent ALMA programs, such as ALPINE and REBELS\footnote{Reionization Era Bright Emission Line Survey \citep{Bouwens_2022_REBELS}.}, have unveiled a population of dusty, obscured star-forming galaxies at $4 \lesssim z \lesssim 9$, which is estimated to contribute $\sim 10-25 \%$ to the $z > 6$ cosmic star formation rate density \citep{Fudamoto_2021}.
The presence of a significant population of red, optically-faint galaxies at these redshifts, especially at the high-mass end of the stellar mass function\footnote{\citet{Gottumukkala_2024} find that the obscured galaxy SMF at $6 < z < 8$ overtakes
the pre-JWST SMF around log$(M_\star/\si{M_\odot}) \sim 10.375$. By integrating the SMF at log$(M_\star/\si{M_\odot}) > 9.25$, they estimate that the stellar mass density might even double with respect to pre-JWST studies.} (SMF), is further confirmed by JWST (e.g. \citealt{Xiao_2023, Gottumukkala_2024}). A lack of prominent emission lines\footnote{Particularly, \citet{Curtis-Lake_2023} reported $2\sigma$ upper limits of $\simeq 6 - 15.4 ~ \si{\angstrom}$ on the HeII equivalent widths. Two out of the four analysed objects also indicate moderate levels of dust (V-band optical depth, $\tau_\mathrm{V} \sim 0.2$), albeit with large uncertainties. However, the study of \citet{DEugenio_2023} demonstrates that deep observations can reveal faint lines that were undetected in shallower spectra, as is the case for GS-z12, one of the galaxies previously analysed in \citet{Curtis-Lake_2023}.} -- possibly ascribed to high levels of dust absorption, and/or a combination of low overall star formation rate and intrinsic faintness -- has also been reported in the spectra of metal-poor galaxies at $z \gtrsim 10$ observed with JWST \citep{Curtis-Lake_2023, Roberts-Borsani_2023}.

A detailed estimate of the dampening of the HeII line caused by interstellar dust would require full radiative-transfer calculations, that are beyond the goals of the present work. However, here we remark that the results shown in Figure~\ref{fig:HeIILuminosityVsRedshift} should be interpreted as an upper limit, while the actual HeII luminosity will depend on the global dust content of the galaxies\footnote{See e.g. \citet{DiCesare_2023} for the dust-to-stellar mass scaling relations of \texttt{dustyGadget} galaxies.} and on their viewing angle, due to their very inhomogeneous dust distribution (\citealt{Venditti_2023}; see also \citealt{Smith_2019_dust}). Focusing on an individual Pop\ III-hosting galaxy at $z = 7.3$, we found that the optical depth ($\tau$) resulting from dust absorption only (i.e. neglecting the contribution of scattering) can vary from $\sim 10^{-8}$ up to $\sim 10$, depending on the specific line-of-sight to the sources \citep{Venditti_2023}. Although particularly unfavourable lines-of-sight might dampen the HeII line by up to a factor $\sim 10^{-9}$ (bringing even the best-case scenario in Figure~\ref{fig:HeIILuminosityVsRedshift} far below the detectability threshold) we note that the $\tau$ distribution is peaked around values of order $\sim 10^{-6} - 10^{-1}$, depending on the position of the considered Pop\ III stellar population relative to the galactic centre (see table~3 and figure~12 of \citealt{Venditti_2023}). Consequently, typical absorption is much lower, up to a factor $\sim 0.9$, i.e. less than 10\% of the line flux is absorbed along typical lines-of-sight. However, we remark that these considerations are based on the study of a single Pop\ III-hosting galaxy, while a more thorough statistical analysis is required to reliably predict the typical dust absorption in such galaxies; moreover, these estimates do not account for the effect of scattering from dust grains.
A strong viewing angle dependence of dust attenuation in high-$z$ galaxies is further demonstrated by \citet{Cochrane_2024}. Although their study specifically focused on a sample of massive and obscured HST-dark galaxies at $4 < z < 7$ rather than Pop\ III-hosting galaxies, this result further supports the notion that predictions of the extinction based solely on the total dust mass are likely insufficient in high-mass galaxies at these redshifts.

\subsection{Detecting Pop\ III through HeII vs. PISNe}
\label{sec:discussion_PISNe}

In \citet{Venditti_2024}, we discussed an alternative channel to identify Pop\ III-hosting galaxies, looking for massive Pop\ III stars ($140 ~ \si{M_\odot} < m < 260 ~ \si{M_\odot}$, \citealt{Heger_Woosley_2002}) at the moment of their death as PISNe. These supernovae are expected to be extremely bright, reaching bolometric luminosities higher than $\sim 10^{45} ~ \si{erg.s^{-1}}$ during the short shock-breakout phase and $\sim 10^{44} ~ \si{erg.s^{-1}}$ during their long-term light-curve evolution \citep{Kasen_2011}, i.e. $\sim 2-3$ orders of magnitude brighter than our most optimistic scenario for the HeII line. Moreover, they could be more straightforwardly identified even without requiring costly spectroscopic analysis\footnote{\citet{Hartwig_2018} and \citet{Moriya_2022_RST} discuss for example the optimal filter combinations to detect PISNe at $z \gtrsim 6$ with JWST and with the Nancy Grace Roman Space Telescope, and to discriminate between different types of supernovae \citep[see also][]{Wang_2012}. In \citet{Venditti_2024} we further discussed how the peak emission from PISNe can easily outshine the stellar emission of their hosting galaxies, especially in spatially resolved observations.}. In fact, the high temperatures required to power HeII line emission can be achieved through a number of other confusing mechanisms/sources, including X-ray binaries \citep{Schaerer_2019, Saxena_2020a, Saxena_2020b, Senchyna_2020, Cameron_2024, Lecroq_2024}, Wolf-Rayet stars \citep{Kehrig_2018, Saxena_2020a, Shirazi_2021, Senchyna_2021, Cameron_2024, Martins_2023, Tozzi_2023, Gomez-Gonzalez_2024}, AGNs \citep{Saxena_2020a, Saxena_2020b, Shirazi_2021, Tozzi_2023, Liu_2024, Topping_2024}, shocks \citep{Kehrig_2018, Lecroq_2024}, and stellar winds \citep{Upadhyaya_2024}.

However, the signal from PISNe is very short-lived, $\sim 1$~yr in the source frame \citep{Kasen_2011}, compared to $\sim 1$~Myr for the HeII line \citep{Schaerer_2002, Schaerer_2003, Katz_2023}. The combination of the short lifetime and the limited mass range of PISNe progenitors makes PISNe extremely rare phaenomena. In \citet[figure~4]{Venditti_2024} we found, at best $\sim 0.4$ PISNe on average among galaxies with $7.5 < \mathrm{log}(M_\star / \si{M_\odot}) \leq 9.5$ at $z \simeq 8$, within the effective volume of all the combined JWST surveys considered in Figure~\ref{fig:HeIIStat_vol}, i.e. more than three orders of magnitude lower than the predicted number of Pop\ III HeII emitters. Hence, a trade-off between the limitation in statistics for PISNe and the limitation in brightness for the HeII signature has to be taken into account when designing our strategies for Pop\ III detection.

\section{Conclusions}
\label{sec:conclusions}

A systematic search for Pop\ III stars during the EoR through the HeII$\lambda$1640 line poses several challenges. We predict more than 400 Pop\ III systems could be discovered in $M_\star \gtrsim 10^{7.5} ~ \si{M_\odot}$ galaxies within existing/ongoing JWST surveys at these redshifts. However, considerable uncertainty surrounds the luminosity of their intrinsic HeII emission, which might vary from $\sim 2 \times 10^{40} ~ \si{erg.s^{-2}}$ up to $\sim 10^{42} ~ \si{erg.s^{-2}}$ depending on the adopted Pop\ III model. The uncertainty is mainly driven by the assumption on the star-formation efficiency parameter $\eta_\mathrm{III}$, while the presence/absence of mass loss only accounts for a factor of order unity. Different assumptions on the Pop\ III IMFs can bring these numbers down by up to a factor $\sim 1/350$. Dust absorption can also further dampen this emission along unfavourable lines-of-sight. While promising candidates such as GN-z11 exist (with an inferred high $\eta_\mathrm{III} \sim 0.1$, and essentially dust-free), it remains unclear how representative such bright targets are. Moreover, many similar targets might fall outside our FoVs, even more than 90 \% when considering small FoVs as for NIRSpec/MOS.

In principle, a large portion of these Pop\ III systems could be too faint to be detected in wide -- but shallow -- blind surveys; for example, the number density $\sim 1 - 2 \times 10^{-4} ~ \si{cMpc³}$ of our Pop\ III systems would be too low to yield realistic detection probabilities in the very deep NIRSpec survey considered by \citet{Vikaeus_2022}, when assuming a low HeII luminosity of $\sim 2.64 \times 10^{40} ~ \si{erg.s^{-1}}$. A more effective strategy might involve follow-up spectroscopy in the regions surrounding bright, massive galaxies: although rarer, these are in fact more likely to host peripheral Pop\ III stars \citep{Yajima_2023, Venditti_2023}. Focusing on a limited number of promising targets would truly allow us to push our instrument capabilities, particularly:
\begin{itemize}
    \item conducting very deep observations, to confirm/exclude the presence of even the faintest Pop\ III systems with high confidence (e.g. an exposure of at least 50~h is required with NIRSpec/IFU at medium resolution to exclude the presence of a $M_\mathrm{III} \gtrsim 2 \times 10^4 ~ \si{M_\odot}$ in a galaxy at $z \sim 6.7$, assuming a Salpeter-like IMF in the range [100, 500]~\si{M_\odot} and strong mass losses; however, the required depth is strongly dependent on the underlying Pop\ III model);
    \item comprehensively sampling the external regions via multiple pointings, to hunt for Pop\ III star-forming clumps in the outskirts (ideally, covering a region up to $\sim 20$~kpc from the galactic center).
\end{itemize}

In \citet{Venditti_2023} we found indications that strong accretion of pristine gas from the IGM at the knots of the cosmic web might favour Pop\ III star formation. On the other hand, under-dense regions with a less progressed history of star formation are also of interest. \citet{CorreaMagnus_2024} suggested a novel formation pathway for Pop IIIs with major mergers as a primary source of gas. However, the role of mergers in the global Pop III star-formation budget -- and hence whether isolated/interacting galaxies are a better observational target -- needs to be confirmed through a more thorough statistical analysis. In future works we plan to delve into all these aspects, to help us identify the most favourable candidates/environments for follow-up.\\

We thank Roberto Maiolino for valuable comments. AV acknowledges support from Sapienza University of Rome program "Bando mobilità internazionale PhD 2022 (II edizione)" (Decreto N. 3147/2022 Prot. n. 0102185 del 15/11/2022) during the visiting period (March-September 2023) at UT Austin, Texas (USA). LG and RS acknowledge support from the PRIN
2022 MUR project 2022CB3PJ3 - First Light And Galaxy aSsembly (FLAGS) funded by the European Union – Next Generation EU, and from the Amaldi
Research Center funded by the MIUR program "Dipartimento di Eccellenza" (CUP:B81I18001170001).

%

\vspace{5mm}
\facilities{
JWST(NIRSpec). Part of the JWST data discussed in this article can be obtained from the Mikulski Archive for Space Telescopes (MAST) at the Space Telescope Science Institute. The specific observations can be accessed via \dataset[doi:10.17909/2dxjz303]{https://doi.org/10.17909/2dxjz303} \citep{Wang_2024}, \dataset[doi:10.17909/2dxjz303]{https://doi.org/10.17909/8tdj-8n28} \citep{Vanzella_2023}.
}


\software{
numpy \citep[\url{https://numpy.org;}][]{VanDerWalt_2011_numpy, Harris_2020_numpy}, 
matplotlib \citep[\url{https://matplotlib.org};][]{Hunter_2007_matplotlib}, 
scipy \citep[\url{https://scipy.org}; \href{https://mail.python.org/pipermail/python-list/2001-August/106419.html}{Jones et al. 2001};][]{Virtanen_2020_scipy}, 
astropy \citep[\url{http://www.astropy.org};][]{AstropyCollaboration_2013, AstropyCollaboration_2018, AstropyCollaboration_2022},
JWST Exposure Time Calculator (\url{https://jwst.etc.stsci.edu}).
}




\bibliography{main}{}
\bibliographystyle{aasjournal}



\end{document}